\documentclass{JHEP3}

\usepackage{epsfig,bbm,bm,amssymb,graphicx}

\author{Subodh P. Patil \\Humboldt Universit\"at zu Berlin, Institut f\"ur Physik, Newtonstra{\ss}e 15, D-12489 Berlin, Germany\\ \\email: subodh@physik.hu-berlin.de\\}

\date{\today}
\preprint{HU-EP-07/60}
\title{Degravitation, Inflation and the Cosmological Constant as an Afterglow}

\newcommand{\eq}[2]{\begin{equation}\label{#1}{#2}\end{equation}}

\abstract{In this report, we adopt the phenomenological approach of taking the degravitation paradigm seriously as a consistent modification of gravity in the IR, and investigate its consequences for various cosmological situations. We motivate degravitation-- where Netwon's constant is promoted to a scale dependent filter function-- as arising from either a small (resonant) mass for the graviton, or as an effect in semi-classical gravity. After addressing how the Bianchi identities are to be satisfied in such a set up, we turn our attention towards the cosmological consequences of degravitation. By considering the example filter function corresponding to a resonantly massive graviton (with a filter scale larger than the present horizon scale), we show that slow roll inflation, hybrid inflation and old inflation remain quantitatively unchanged. We also find that the degravitation mechanism inherits a memory of past energy densities in the present epoch in such a way that is likely significant for present cosmological evolution. For example, if the universe underwent inflation in the past due to it having tunneled out of some false vacuum, we find that degravitation implies a remnant `afterglow' cosmological constant, whose scale immediately afterwards is parametrically suppressed by the filter scale ($L$) in Planck units $\Lambda \sim l^2_{pl}/L^2$. We discuss circumstances through which this scenario reasonably yields the presently observed value for $\Lambda \sim O(10^{-120})$. We also find that in a universe still currently trapped in some false vacuum state, resonance graviton models of degravitation only degravitate initially Planck or GUT scale energy densities down to the presently observed value over timescales comparable to the filter scale. We argue that different functional forms for the filter function will yield similar conclusions. In this way, we argue that although the degravitation models we study have the potential to explain why the cosmological constant is not large in addition to why it is not zero, it does not satisfactorily address the co-incidence problem without additional tuning.\footnote{Dedicated to the memory of Kris Bellamkonda}}

\begin{document}


\section{Introduction}
The degravitation paradigm \cite{addg}\cite{dhk}\cite{new} is a phenomenological proposal designed to solve the cosmological constant (CC) problem directly. Rather than positing some symmetry which sets the CC to be vanishing, or invoking anthropic arguments such that the problem is paraphrased away, the degravitation paradigm invokes an IR modification to general relativity at the level of the equations of motion, such that Newton's constant becomes dependent on the wavelength of the source. Specifically, taking $\lambda$ as a characteristic length scale of some source, and taking $L$ to be some IR length scale to be specified later, degravitation requires that Newton's constant be promoted to some function $G_N(\lambda, L)$ such that the following properties are satisfied:

\begin{eqnarray}
\label{sd}
G_N(\lambda,L) \to &0&~~~  \lambda \gg L,\\
\nonumber G_N(\lambda,L) \to &G_N&~~~  \lambda \ll L.
\end{eqnarray}

\noindent In this way we see that sources with wavelengths greater than the IR scale introduced by $L$ effectively degravitate as their gravitational coupling vanishes, whereas sources with wavelengths less than $L$ gravitate normally \footnote{Requiring that gravity be unaffected at presently accessible scales requires us to set this IR scale to be larger than the present horizon scale.}. Such scale dependent coupling, although seemingly odd in the context of gravity, has many precedents in gauge theory. As example from electromagnetism \cite{addg}, we note that in a linear dielectric medium, the electric field responds to a source as

\eq{es}{\nabla \cdot E = \frac{\rho}{4\pi\bar{\epsilon}}~~~~\bar{\epsilon} = \epsilon_0(1 + \kappa),}

\noindent where $\kappa$ is the dielectric constant of the ambient medium. For a frequency dependent dielectric, we see that if the medium is such that $\bar\epsilon \gg 1$ for sources with characteristic frequency $\omega \gg 1/T$ (where $T$ is some time scale), then all such sources will effectively 'de-electrify' and not source the electric field. This is effectively the underlying physics of a high pass filter.

\par
As proposed in \cite{addg} and further developed in \cite{dhk}\cite{new}, a scale dependent in Newton's constant (\ref{sd}) can be implemented by promoting it to a differential operator:

\eq{gn}{G_N \to G_N(L^2\square),}

\noindent where $\square$ is the covariant D'Alembertian operator, and $L$ is taken to be a large but finite IR cutoff scale, typically much larger than any scale presently accessible to cosmology. As a self-adjoint differential operator, $\square$ is guaranteed to have a complete set of mode functions with real eigenvalues. Expanding any source in terms of these mode functions, we see that mode by mode, the effect of acting on a source with the operator (\ref{gn}) is to make the replacement

\eq{gna}{G_N \to G_N(L^2 \lambda^{-2}),}

\noindent where $\lambda^{-2}$ is the eigenvalue of the corresponding mode function being acted on. If for heuristic purposes we were to consider the example of Minkowski space, then the mode functions of the D'Alembertian are plane waves, and the effect of acting on a mode described by the vector $k^\mu$, is that Newtons constant becomes the scale dependent function $G_N(L^2k^2)$. Hence degravitating all sources with wavelengths larger than $L$ requires that

\eq{ee}{G_N(\alpha) \to 0~~ \alpha \ll 1.}

\noindent As a source which is constant in both space and time, the cosmological constant corresponds to the zero mode $k^\mu = 0$, with the corresponding eigenvalue $\lambda^{-2} = 0$, and is hence annihilated by the action of the differential operator (\ref{gn}) by virtue of (\ref{ee}). In this way, we see that any source of the form

\eq{ccs}{T^\mu_\nu = \Lambda \delta^\mu_\nu}

\noindent effectively degravitates. At this rather heuristic level, it is trivial to observe that degravitation kills more than just the spacetime zero mode, as plane wave sources (also with eigenvalues $\lambda^{-2} = k^2 = 0$) degravitate as well. In fact, in a general background, any source with an energy density which lies in the null eigenspace of the D'Alembertian degravitates. For the purposes of cosmology, this is only of passing interest as most cosmological sources other than the cosmological constant gravitate normally \footnote{It is straightforward to show that although plane waves degravitate, radiation gases gravitate normally due to decoherent superposition, and source the usual FRW radiation dominated expansion.}. It remains then to realize (\ref{gn}) from some underlying physical model. There are several distinct ways through which this might occur, and we discuss them presently.

\subsection{Filtering via a resonantly massive graviton}

In \cite{dhk}, and further developed in \cite{new}, a class of models in which degravitation is effected by massive or resonance gravitons was shown to result in a filter function for Newton's constant of the form:

\eq{rg}{8\pi G_N \to \frac{8\pi G_N}{1 + (\frac{m^2}{\square})^{1-\alpha}}, ~~0\leq \alpha < 1,}

\noindent where $\alpha = 0$ corresponds to a theory of massive gravity, and $\alpha = 1/2$ corresponds to a filter function arising in DGP braneworld constructions \cite{dgp} \footnote{$G_N$ is the usual 4-dimensional Newtons constant in the above.}. One obtains such a filter function by the argument provided in \cite{dhk}, which we sketch here. We begin with the Einstein equations, linearized around Minkowski space:

\eq{ehprop}{\Omega_{\mu\nu}^{\alpha\beta}h_{\alpha\beta} := \square h_{\mu\nu} - \eta_{\mu\nu}\square h - \partial_\mu\partial^\alpha h_{\alpha\nu} - \partial_\nu\partial^\alpha h_{\alpha\mu} + \eta_{\mu\nu}\partial^{\alpha}\partial^\beta h_{\alpha\beta} + \partial_\mu\partial_\nu h = -8\pi G_N T_{\mu\nu}.}
By adding the Fierz-Pauli mass term \cite{fp}, which is the only ghost free form for the propagator of a massive spin two field, we obtain the equation of motion:
\eq{fpprop}{\Omega_{\mu\nu}^{\alpha\beta}h_{\alpha\beta} -m^2[h_{\mu\nu} - \eta_{\mu\nu} h] = -8\pi G_N T_{\mu\nu}.}
Group theoretic considerations tell us that a massive spin two particle has five polarizations, as opposed to the usual two for a massless spin two particle. As a way of book keeping the extra polarizations, we perform the St\"uckelberg decomposition, and write $h_{\mu\nu}$ as
\eq{stuckd}{h_{\mu\nu} = \tilde h_{\mu\nu} + \partial_\mu A_\nu + \partial_\nu A_\mu,}
with $\tilde h_{\mu\nu}$ being the helicity two piece, and with $A_\mu$ carrying the extra polarizations. We note that this decomposition introduces the following gauge symmetry:

\eq{gss}{ \tilde h_{\mu\nu} \to \tilde h_{\mu\nu}+ \partial_\mu \xi_\nu + \partial_\nu \xi_\mu,~~ A_\mu \to A_\mu - \xi_\mu.}
Taking the divergence of (\ref{fpprop}) once we have St\"uckelberg decomposed $h_{\mu\nu}$, we end up with the equation of motion for $A_\mu$:
\eq{ameq}{\partial^\mu F_{\mu\nu} = -\partial^\mu\Bigl(\tilde h_{\mu\nu} - \eta_{\mu\nu} \tilde h \Bigr),}
with $F_{\mu\nu} = \partial_\mu A_\nu - \partial_\nu A_\mu$. Solving for $A_\mu$ using this equation, and substituting this back into (\ref{fpprop}), one can show after accounting for residual gauge invariances \cite{dhk} that the resulting equations for the helicity two component is

\eq{hel2}{\Bigl(1 + \frac{m^2}{\square} \Bigr)\Omega_{\mu\nu}^{\alpha\beta}\tilde h_{\alpha \beta} = - 8\pi G_N T_{\mu\nu}.}
One can generalize the Fierz-Pauli mass term to allow for a resonance graviton, where $m^2 \to m^2(\square)$, such that  
\eq{hel2.2}{\Bigl(1 + \frac{m^2(\square)}{\square} \Bigr)\Omega_{\mu\nu}^{\alpha\beta}\tilde h_{\alpha \beta} = -8\pi G_N T_{\mu\nu}}
results. This is the basis for proposing (\ref{rg}) as filter function for Netwon's constant. However, there are some immediate caveats to take note of here. Certainly (\ref{hel2}) was derived around a linearized approximation around Minkowski space. Although massive gravity can be defined around other backgrounds \cite{des}, one can only make sense of it in the linearized approximation. In fact, its non-linear generalization is problematic and results in the instability of Minkowski space \cite{gg}. Hence to take (\ref{rg}) as it stands literally would be a mistake, not least because in addition, satisfying the Bianchi identities would impose new constraints on top of the usual covariant conservation (used to derive (\ref{ameq})). Although the spirit of the investigation that follows is to only consider filter functions of the form (\ref{rg}) phenomenologically, such an investigation would only be meaningful if our formalism is self-consistent. Hence it is neccesary to address the issue of the non-linear completion of (\ref{rg}) as well as the issue of satisfying the Bianchi identities. As we demonstrate next, there is another possibility in realizing a non-trivial filtering of gravity as in (\ref{gn}) which addresses both of these anxieties.

\subsection{Filtering through running of $G_N$}

The running of coupling constants is a well understood phenomenon in quantum field theory. The nature of the running is determined by the renormalization group equations (RGE) of the theory in consideration. For a theory with a single coupling parameter, such as quantum electrodynamics, the RGE takes the form

\eq{rge}{\mu \frac{d \alpha}{d \mu} = \beta(\alpha),}
where $\mu$ is the energy scale of interest (the external momenta flowing into a given process). In a typical gauge theory, solving for the differential equation that is the RGE results in the following expression for the running couplings:

\eq{rc}{ \alpha_i(\mu) = \alpha_i(\mu_0) + c_i ln~(\mu/\mu_0),}
where $\mu_0$ is some reference energy scale where the coupling strength is experimentally measured, and $c_i$ is determined by the RGE at hand. When computing physical quantities derived from scattering amplitudes, we usually work in momentum space. However, if for some reason we wanted to implement the running of couplings in configuration space, we could do so via the replacement \cite{barv1}\cite{barv2}\cite{spectral}: 

\eq{abox}{\alpha \to \alpha(\square) = \alpha_0(\mu_0) + c ~ln[\square/\mu_0^2],}  
where $\mu_0$ is some reference energy scale such that $\alpha(\mu_0) = \alpha_0$, and $c$ depends on the RGE of the theory in question. A simple context in which one can illustrate this explicitly is furnished by QED, where the interaction Lagrangian is given by:
\eq{lint}{\mathcal L_{int } = e\bar\Psi \gamma^\mu A_\mu \Psi(x).}
In Fourier space, the net effect of running our coupling, is that physical processes become functions of the scale of the process at hand, set by the external momenta $p$, i.e.: $e\to e(p^2/\mu_0^2)$. From the properties of Fourier transforms, it is easy to see that taking the coefficient of a product of local operators to now also depend on momentum, `de-localizes' the vertex. This is captured by the prescription of rewriting the coupling as $e(\square/\mu^2_0)$, where the delta function constraint on the vertex effects the replacement $\square \to p^2$. This is to say, in position space, (\ref{abox}) encodes (typically non-local) quantum corrections to the classical equations of motion. We illustrate this in context. In QED, we know that corrections to the photon propagator from virtual fermion loops results in the running 
\eq{ferml}{\alpha \to \alpha(k^2) = \alpha\Bigl[1 + \frac{\alpha}{3\pi}ln~\frac{-k^2}{m_e^2} + O(e^4_0)\Bigr] ~~; k^2/m_e^2 >> 1,}   
with $\alpha \approx 1/137$ being the low energy fine structure constant. In configuration space, this results in the replacement
\eq{fermlnl}{\alpha \to \alpha\Bigl[1 + \frac{\alpha}{3\pi}ln~[-\square/m_e^2] + O(e^4_0)\Bigr],}
which effects the observed quantum corrections to the classical equations of motion. For example, we know that the electrostatic potential between an electron and an infinitely heavy point charge (with momentum transfer $k^\mu = (0,\vec k)$) is obtained from the inverse Fourier transform of the scattering amplitude:
\eq{pot}{V(r) = \int \frac{d^3k}{(2\pi)^3} e^{i\vec k\cdot \vec r}\frac{4\pi\alpha(- k^2)}{k^2}.}
Taking the fine structure constant to be truly constant ($4\pi\alpha = e_0^2$) results in the usual classical electrostatic potential:
\eq{classans}{V(r) = \frac{e_0^2}{4\pi r}.}
However, we know that the coupling does in fact run, and the net effect of this is that quantum effects effectively smear out the charge into a charge distribution given by \cite{huang}
\eq{ecd}{\rho(r) = \int \frac{d^3k}{(2\pi)^3} e^{i\vec k\cdot \vec r}4\pi\alpha(- k^2),}
which can be implemented in (\ref{pot}) through the prescription (\ref{abox}):

\eq{nlpot}{V(r) = \int \frac{d^3k}{(2\pi)^3} 4\pi\alpha(- \square) e^{i\vec k\cdot \vec r}\frac{1}{k^2}.}
In the limit of a vanishing electron mass ($m_er \ll 1$ so we can use (\ref{ferml}) in the above), we find the quantum corrected potential to be
\eq{qcpot}{V(r) =  \frac{e_0^2}{4\pi r}\Bigl[1 + \frac{\alpha}{3\pi}\Bigl( ln\frac{1}{m_e^2r^2} -2\gamma \Bigr)\Bigr],}
with $\gamma$ being the Euler constant ($-2\gamma \approx -1.154$). Hence we see how running couplings in gauge theories effects modifications to the classical equations of motion, such that it manifests as if the coupling constants were promoted to functions of $\square$. We now ask, is this possible in gravity? We begin with a suggestive toy example.

Quantum gravity in two dimensions is a renormalizable theory. This is due to the fact that Newton's constant in 2-d is dimensionless. We could couple gravity to any number of matter and gauge fields, and compute the one loop beta function for Newton's constant in such a setting, as was done by Christensen and Duff in \cite{duff}. There one finds (working in $d = 2 + \epsilon$ dimensions)

\eq{bgrav}{\beta(G) = (d-2)G - \beta_0G^2 - ...,}
where $\beta_0$ can be computed as
\eq{b0}{\beta_0 = \frac{2}{3}[1 - n_{3/2} + n_{1/2} - n_0 - N_1 + N_{1/2} - N_0],}
where our field content consists of $n_s$ massless spin s particles and $N_s$ massive spin s particles. From (\ref{bgrav}) we see immediately the celebrated UV fixed point for gravity in $2 + \epsilon$ dimensions at $G_c = (d-2)/\beta_0$, for positive $\beta_0$. The differential equation defining the beta function:
\eq{diffeq}{\mu\frac{dG(\mu)}{d\mu} = \beta(G)}
can be solved to give us the running of $G$:
\eq{runga}{G(k^2) = \frac{G_c}{1 \pm (m^2/k^2)^{\frac{d-2}{2}}},} 
where $m^2$ is now an integration constant which corresponds to the scale where we experimentally determine $G$. In configuration space, this corresponds to the filter function
\eq{ff}{G(\square) = \frac{G_c}{1 \pm (m^2/\square)^{\frac{d-2}{2}}},}
which is to be compared to the one parameter family of filter functions proposed in \cite{dhk} (\ref{rg}):
\eq{mgff}{G(\square) = \frac{G_c}{1 + (m^2/\square)^{1-\alpha}},}
where $0 \leq \alpha < 1$. The question of how one would generalize this to four dimensions to obtain a suitably degravitating filter function for $G_N$ is presently the subject of an investigation \cite{ghp}. For the present purposes however, we only wish to draw from this example the lesson that the effects of a running $G_N$ would be to promote it to a function of $\square$ at the level of the field equations. Although the precise functional form might differ from (\ref{rg}), the objectives of degravitation will be met provided the condition (\ref{ee}) is satisfied. In fact, although we shall stick to (\ref{rg}) in much of what follows as a phenomenological example through which to study degravitation, we argue further on by example that the conclusions we present in what follows depends only on the filter scale $L$ ( $= m^{-1}$ in the above), and not the precise functional form of the filter. In addition, as we shall see in the next section, understanding degravitation as a semi-classical effect also allows us new perspectives on how the Bianchi identities are to be satisfied. We now outline the investigation to follow.

\subsection{Outline}

In this report, we study the various cosmological consequences of promoting Newton's constant to an operator valued filter function. We adopt a phenomenological approach in this report, and take (\ref{rg}) as an example filter function which might have arisen from any number of ways-- integrating out extra polarizations to the graviton arising from either a small mass or through extra dimensional effects, or from the running of Newton's constant as illustrated above. We argue further that the conclusions we draw from this example filter function generalize to other functional forms of suitably degravitating filter functions.

We begin by setting up the formalism and addressing the issue of defining and solving the modified Einstein's equations with a filter function for Newton's constant, taking care to satisfy the Bianchi identities in this context. We then proceed to consider our sample filter in the context of various cosmological scenarios, namely slow roll inflation, hybrid inflation and old inflation, and show that in spite of neutralizing the cosmological constant, degravitation leaves these scenarios qualitatively unchanged. We then uncover the result that in a universe which is the end result of having tunneled out of some false vacuum (vacua) in the past, there exists a present day `after-glow' cosmological constant whose scale is parametrically suppressed by the square of the filter scale in Planck units. Given that we should set the filter scale such that it should be larger than the present day horizon scale so as not to conflict with cosmological observations, we find $\Lambda \sim l^2_{pl}/L^2 < l^2_{pl}H^2_0 \sim 10^{-120}$ in reduced Planck units. We then uncover the result that in the scenario where our universe remains stuck in some false vacuum state, degravitation only suppresses this false vacuum density enough to yield a cosmological constant similar to the one observed today with additional fine tuning. We discuss this last result in the context of the coincidence problem, after which we offer our conclusions. In the following we work in natural units, and unless otherwise states, take our metric tensor to be

\eq{mt}{diag[1,-a(t)^2,-a(t)^2,-a(t)^2].}

\noindent The 4-dimensional Planck mass is taken to be $M^2_{pl} = (8\pi G_N)^{-1}$, where $G_N$ is the 4-dimensional Newton's constant.

\section{The Bianchi Identities and the Modified Einstein Equations}
As first proposed in \cite{addg}, we consider the following formal modification of Einstein's equations:

\eq{med}{G^\mu_\nu = 8\pi G_N(L^2\square) T^\mu_\nu.}
One encounters the immediate curiosity that since in general $[\nabla_\mu,\square] \neq 0$, either the Bianchi identities are not satisfied in the above, or new constraints are imposed on a covariantly conserved energy momentum tensor. We wish to demonstrate that it is in fact the latter that is true, and that the correct Bianchi identity is indeed $\nabla_\mu [G(L^2\square)T^\mu_\nu] = 0$. 

In the context of obtaining a filtering $G_N$ as the configuration space expression of its running, this is to be expected considering the fact that in reality, the equations (\ref{med}) are actually the equations
\eq{semi}{G^\mu_\nu = \frac{1}{M^2_{pl}}\langle T^\mu_\nu \rangle}
in disguise. As we encountered with the previous examples, just as quantum effects modify the classical equations of motion so as to introduce non-local operators (\ref{abox}) in place of coupling constants (so that point sources get smeared out (\ref{ecd})), so we expect the quantum mechanical running of $G_N$ to give us (\ref{med}) as an paraphrasal of (\ref{semi}), the right hand side of which is to be obtained from the effective action as:
\eq{effact}{\langle T_{\mu\nu} \rangle = \frac{2}{\sqrt{-g}}\frac{\delta W}{\delta g^{\mu\nu}}.}
We observe that the effective Lagrangian density (we take for illustrative purposes, a massive degree of freedom with mass $m$) is given by the expression \cite{bd}:
\begin{eqnarray}
\label{effld} W &=& \int d^4x \sqrt{-g(x)}~\mathcal L_{eff}(x)\\ 
 \mathcal L_{eff} &=& \frac{i}{2}\lim_{x'\to x}\int^\infty_{m^2}d\bar m^2 G_F^{DS}(x,x';\bar m^2),
\end{eqnarray}
where $G^F_{DS}(x,x';\bar m^2)$ is the Feynman propagator for a field with mass $\bar m$ in the Schwinger De-Witt representation. We note that it is the part of $\mathcal L_{eff}$ which remains finite in the limit $x'\to x$ which is sensitive to the large scale structure of the manifold (i.e. is non-local) and to the quantum state defining the expectation value \cite{bd}.

In this way, it should not be surprising that the true quantum corrected equations of motion should be $\nabla_\mu [G(L^2\square)T^\mu_\nu] = 0$, as $\nabla_\mu T^\mu_\nu = 0$ implies only the classical equations of motion, derived from the classical Lagrangian density $\mathcal L_{c}$ as opposed to $\mathcal L_{eff}$. Thus we understand the modified Bianchi identities as actually encoding quantum corrections to our classical equations of motion. This can also be understood as coming from the non-commutativity of taking the expectation value of an operator, and taking its covariant derivative one we account for the fact that the perturbative vacuum differs from the free vacuum. We flesh this argument out with the example of a free scalar field, which has the energy momentum tensor:
\eq{emtsf}{T_{\mu\nu} = \nabla_\mu\phi\nabla_\nu\phi - \frac{1}{2}g_{\mu\nu}\nabla_\kappa\phi\nabla^\kappa\phi + \frac{m^2}{2}g_{\mu\nu}\phi^2.}
We define the above bilinear quantity as $T_{\mu\nu} = T_{\mu\nu}[\phi,\phi]$. If the field admits the mode expansion
\eq{modex}{\phi(x,t) = \int d^3k~\Bigl( u_{\vec k}a_{\vec k} + u^{*}_{\vec k}a^{\dagger}_{\vec k}\Bigr),}  
where $u_k$ is the appropriate mode function, indexed by $\vec k$ such that the vacuum is defined as $a_{\vec k}|0\rangle = 0 ~\forall \vec k$. It is easy to check \cite{bd} that
\eq{vev}{\langle 0|T_{\mu\nu}|0\rangle = \int d^3k ~ T_{\mu\nu}[u_{\vec k},u^{*}_{\vec k}],}
whereas for non vacuum expectation values:
\eq{nonvev}{\langle n_1(k_1),n_2(k_2)... |T_{\mu\nu}| n_1(k_1),n_2(k_2)... \rangle = \int d^3k ~ T_{\mu\nu}[u_{\vec k},u^{*}_{\vec k}] + 2\int d^3k ~ \sum_i n_i(k) T_{\mu\nu}[u_{\vec k},u^{*}_{\vec k}].} 
We note from the form of (\ref{emtsf}), that the graviton propagator will receive quantum corrections from scalar field loops. In addition, any scalar self couplings will also run, and the vacuum of the interacting theory will differ from the vacuum of the free theory. We can understand how quantum effects alter the classical equations of motion by the following argument. Although it is clear from (\ref{vev}) that $\nabla^\mu\langle 0|T_{\mu\nu}|0\rangle = 0$ (as on the right hand side we simply have the mode decomposition of $\nabla^\mu T_{\mu\nu} = 0$, where by construction, the mode functions satisfy the equations of motion), it should also be clear that the presence of non zero occupation numbers $n_i(k)$ makes the inverse transform of the divergence of the second term on the right hand side of (\ref{nonvev}) no longer vanish. Considering that the vacuum defined by the mode functions used above is not the true perturbative vacuum for the interacting theory, the interacting vacuum expectation value $\langle 0_I| T_{\mu\nu} |0_I \rangle$ is in general going to contain non-zero particle occupation numbers of the free vacuum (i.e. terms of the form (\ref{nonvev}) will result). Hence non-trivial vacuum structure results in non-satisfaction of the classical Bianchi identities as a result of (\ref{nonvev}), which implies quantum corrected equations of motion. Thus the result of inverting the mode expansion of the vev of the energy momentum tensor in the interacting vacuum back to configuration space (in flat space, we'd simply Fourier transform), is that we would obtain the modified equation of motion (\ref{mee}):
\eq{degmod}{G^\mu_\nu = \frac{1}{M^2_{pl}}\langle 0_I| T_{\mu\nu} |0_I \rangle = 8\pi G_N(\square) T^\mu_{\nu(0)}}
where $T^\mu_{\nu(0)}$ is the classical energy momentum tensor corresponding to the non-interacting vacuum.

In the event that our gravitational filter arises from a small resonant mass to the graviton (perhaps through extra dimensional effects), the new constraints imposed by the Bianchi identities are to be understood through the following argument proposed in \cite{dhk}. The filter function in (\ref{med}) is obtained through integrating out the extra polarizations of a massive spin two theory, which contains five propagating degrees of freedom as opposed to the usual two of Einstein gravity. However the Einstein tensor in (\ref{med}) is constructed only from the spin two sector of this theory, therefore the energy momentum tensor should not be conserved with respect to covariant differentiation with the spin two metric, as it is only conserved with respect to the full metric which contains the extra polarizations. In this way, the non-conservation of the right hand side of the above is accounted for by the ever present extra polarizations \cite{dhk}.
With this in mind, we proceed to study the cosmological consequences of the modified Einstein equations, to be viewed as the phenomenological manifestation of whatever the physics might be that underlies the filtering of gravity. 

Before we continue however, we wish to note that in general, the action of $\square$ on a rank two tensor mixes up its components non-trivially. However, since the covariant d'Alembertian always commutes with the metric tensor, we can avoid this concern by simply considering the trace of the above:

\eq{tr}{R = - 8\pi G_{N}(L^2\square)T,}
In this form, we do not even have to make a specific ansatz for the form of the metric tensor to write down a formal expression for curvature as a function of a matter source.





Having set up the bare basics of the degravitation framework, we turn our attention towards inflationary cosmology in this setup. As articulated in \cite{rhb}, a long standing concern directed at any proposed direct solution to the cosmological constant problem is that it should not also render inflation null. Since inflation depends on the time independent piece of a scalar field potential, this is a valid concern and as we will shortly see, one that the degravitation paradigm answers in a rather appealing manner.

\section{Inflation (v.s. the Cosmological Constant)}

Consider a universe that is dominated by matter with an equation of state parameter $w = -1$.
\eq{mee}{8\pi G_N(L^2\square)\rho = 8\pi G_N(L^2\square)V(\phi).}

\noindent Focussing on a filter function of the form:

\eq{mg}{8\pi G_N(L^2\square) = \frac{8\pi G_N}{1 + \frac{m^2}{\square}},}

\noindent we can immediately prove that such a filter function does in fact degravitate sources with wavelengths larger than the filter scale. Consider a mode expansion of source in terms of homogeneous plane waves\footnote{Although these are not technically the eigenfunctions of the D'Alembertian in a de Sitter spacetime (which are hypergeometric functions), for homogeneous sources they do satisfy $\square e^{i\omega t} = \lambda e^{i\omega t}$, where $\lambda = -\omega^2 + 3iH\omega$ is a complex `eigenvalue'.}, and consider the action of (\ref{mg}) on such a mode. Were we to expand (\ref{mg}) in terms of its power series, we have to take care of the fact that we must expand within the radius of convergence. That is, if $\square$ acts on such a mode function with eigenvalue $-k^2$, then we have two possible expansions:

\begin{eqnarray}
\label{be}
\frac{1}{1 + \frac{m^2}{\square}} &=& \sum_{n=0} \Bigl(\frac{-m^2}{\square}\Bigr)^n  =  \sum_{n=0} \Bigl(\frac{m^2}{k^2}\Bigr)^n ~~;~~m^2 < k^2\\
\label{se}
\frac{1}{1 + \frac{m^2}{\square}} &=& \frac{\square}{m^2}\frac{1}{1 + \frac{\square}{m^2}} = \sum_{n=0} \Bigl(\frac{-\square}{m^2}\Bigr)^n\frac{\square}{m^2}  =  \sum_{n=1} -\Bigl(\frac{k^2}{m^2}\Bigr)^n ~~;~~k^2 < m^2.
\end{eqnarray}

\noindent Where we draw attention to the fact that the sum in (\ref{se}) commences at $n=1$ whereas the sum in (\ref{be}) commences at $n = 0$. From all this we see that for all sources with wavelengths much greater than the filter scale ($L = m^{-1}$) effectively degravitate:

\eq{db}{G_{eff} = G_N O(k^2/m^2)  ~~ k^2 \ll m^2,}

\noindent whereas all sources with wavelengths much smaller than the filter scale gravitate almost normally:

\eq{ds}{G_{eff} = G_N(1 + O(m^2/k^2))  ~~ m^2 \ll k^2.}

\noindent In other words, because the leading order term in (\ref{se}) is proportional to $\square$, any bona fide spacetime zero mode is annihilated by the degravitation filter. One might then wonder how this is consistent with causality, for how is the degravitation filter to know immediately whether an energy density is a legitimate zero mode, or whether it might change in the future? The answer to this lies in the fact that the filter is in fact sensitive to local changes in the energy density, as evidenced by the continuous dependence of (\ref{se}) on the wavenumber of the source. Any local variation, no matter how small will result in a non-zero Newton's constant in a manner that continuously approaches zero for a spacetime zero mode. In fact, as we shall see shortly, if our energy density were to suddenly jump from one constant in the present to another in the future, then the filter does not annihilate the source, except in the far future once the source has had enough time to look like a zero mode to an asymptotic observer.

Since we are interested in cosmological applications, we first restrict ourselves to the effect of the degravitation filter on homogeneous sources (with respect to the spatial coordinates). We can express such a source in terms of its inverse Laplace transform:

\eq{ilt}{\rho (t) = \frac{1}{2\pi}\int^\infty_{-\infty} d\omega e^{i\omega t} \rho(i\omega),}

\noindent where the Laplace transform is defined as

\begin{eqnarray*}
\label{lt}
\rho(\omega) &=& \int^\infty_0 dt' e^{-\omega t'}\rho(t'),\\
\nonumber
&=& \int^\infty_{-\infty} dt' e^{-\omega t'}\tilde\rho(t'),
\end{eqnarray*}

\noindent with the definition

\begin{eqnarray*}
\label{ltdef}
\tilde \rho (t) &=& \rho (t) ~~~~ t > 0\\
&=& 0~~~~~~~~ t < 0.
\end{eqnarray*}

\noindent The use of the Laplace transform is relevant for a universe with a beginning in the finite past. To model an eternal universe we simply set $\tilde \rho(t) = \rho(t)$ for all times and understand the above as a Fourier transform. We now consider the effect of the filter (\ref{mg}) on a homogeneous source
\footnote{For the moment, we make the approximation that in an inflationary universe, the time scale set by the filter scale $\tau = m^{-1}$ is far greater than at least one scale in the problem (e.g. the Hubble scale $H^{-1}$, or the time scale over which we are considering cosmological evolution). We will attempt to relax this requirement further on, but for the time being it allows us to simply act through with the D'Alembertian on (\ref{degs}), as it ensures that any time dependence in $H$ (inferred from the $00$ modified Einstein equation) will be negligible. This is certainly justified up to terms of order $m^4$.}:

\eq{degs}{\frac{8\pi G_N}{1 + \frac{m^2}{\square}}\rho (t) = 8\pi G_N\int^\infty_{-\infty} d\omega \frac{e^{i\omega t} (\omega^2 - 3iH\omega)}{\omega^2 - 3iH\omega - m^2}\frac{\rho(i\omega)}{2\pi}.}

\noindent In order to study the effect of degravitation on various inflationary scenarios, it suffices to consider energy densities corresponding to two special cases-- a slowly varying potential, and a potential which is constant everywhere except for a finite jump/ drop at a specific time (i.e. a step function).

In frequency space, a slow rolling potential can be modeled as the real part of

\eq{vo}{V(\omega) = V_0\delta(\omega) + V_1\delta(\omega - \epsilon),}

\noindent which corresponds to the potential

\eq{pots}{V(t) = V_0 + V_1e^{i\epsilon t},}

\noindent where $\epsilon$ is taken to be some very small frequency (relative to the degravitation and Hubble scales). Evaluating (\ref{degs}) in this context is rather straight forward:

\eq{dv}{V_{deg}(t) = V_0 \tilde\delta(m^2) + V_1\frac{e^{i\epsilon t}}{1 - \frac{m^2}{\epsilon^2 - 3iH\epsilon}},}

\noindent where $V_{deg}(t)$ is the degravitated potential, and $\tilde\delta(m^2)$ is formally defined as

\begin{eqnarray}
\label{tild}
\tilde\delta(m^2) &=& 1 ~~,~~m^2 = 0\\
\nonumber &=& 0 ~~,~~ m^2 \neq 0.
\end{eqnarray}

\noindent In this way, we again see how degravitation immediately annihilates any pure spacetime zero mode for any finite filter scale. For the slowly varying piece, for $\epsilon \gg m$ and $H \gg m$ (which corresponds to the field velocity and the Hubble scales being much less than the time scale associated with the filter scale), we see then that (\ref{dv}) reduces to (\ref{pots}) immediately. In fact for chaotic slow roll inflation, we find that $\epsilon$ is typically given by

\eq{ep}{\epsilon \sim m_\phi/\tilde \epsilon,}

\noindent where $m_\phi$ is the mass of the inflaton and $\tilde\epsilon$ is the standard dimensionless slow roll parameter. From this, we see that the condition $m/\epsilon = \tilde\epsilon m/m_\phi\ll 1$ is very easy to satisfy given that the filter scale is to be larger than the present Hubble scale in order not to effect gravity at presently accessible scales ($m^{-1} > H^{-1}_0$), and that the mass of the inflaton is bound to be tens of orders of magnitude greater than this. Thus we conclude that aside from the non contribution of the zero mode (which is typically set to zero by hand anyway), slow roll inflation is virtually unaffected by the degravitation filter once we account for the relevant scales.

The next example we turn to is that of a step function (in time) change in energy density, which is of interest to us for its utility in modeling various scenarios of interest. We represent the step function as

\eq{hsf}{\theta(t-t_0) = \int^\infty_{-\infty}d\omega ~e^{i\omega(t-t_0)}\Bigl[\frac{\delta(\omega)}{2} + \frac{1}{2\pi i\omega}\Bigr].}

\noindent By applying the degravitation filter to the above, we obtain

\eq{dgsf}{\Bigl(1 + \frac{m^2}{\square}\Bigr)^{-1}\theta(t - t_0) = \int^\infty_{-\infty} d\omega \frac{e^{i\omega (t - t_0)} (\omega^2 - 3iH\omega)}{\omega^2 - 3iH\omega - m^2} \Bigl[\frac{\delta(\omega)}{2} + \frac{1}{2\pi i\omega}\Bigr],}

\noindent which evaluates to

\eq{dgsfp}{\Bigl(1 + \frac{m^2}{\square}\Bigr)^{-1}\theta(t - t_0) = \frac{\tilde\delta(m^2)}{2} + \frac{1}{2\pi i}\int^\infty_{-\infty} d\omega \frac{e^{i\omega (t - t_0)} (\omega - 3iH)}{(\omega - \omega_+)(\omega - \omega_-)},}

\noindent with $\tilde\delta(m^2)$ defined in (\ref{tild}), and where the denominator factors into the roots

\eq{roots}{\omega_\pm = \frac{3iH}{2}\Bigl(1 \pm \sqrt{1 - \frac{4m^2}{9H^2}}\Bigr).}

\noindent we see that (\ref{dgsfp}) is readily evaluated in light of the pole structure at the points $\omega_\pm$. We see that for non zero values of $m$, both poles lie above the real axis, and when $m$ vanishes, $\omega_-$ lies on the real axis. Hence for $t < t_0$, in closing the contour in (\ref{dgsfp}) in the lower half plane, we encounter no poles unless $m^2 = 0$, in which case we get a contribution of $-1/2$. That is, for $t < t_0$ we get

\eq{dgb}{\Bigl(1 + \frac{m^2}{\square}\Bigr)^{-1}\theta(t - t_0) = \frac{\tilde\delta(m^2)}{2} - \frac{\tilde\delta(m^2)}{2} = 0~~,~~ t < t_0}

\noindent which is exactly what we would have got in the undegravitated case. This should not be surprising, as degravitation is engineered to preserve causality, and up to $t < t_0$ should not be able to tell the difference between a step function (with the step yet to come), and a zero mode. Proceeding similarly for $t > t_0$, we find in closing the contour in the upper half plane we will always enclose both roots, except that in the case where $m^2 = 0$, the root at $\omega_-$ contributes half the residue it would otherwise as it then lies on the real axis rather than above it. With this in mind, we can evaluate (\ref{dgsfp}) for $t > t_0$ as

\begin{eqnarray}
\label{dga}
\Bigl(1 + \frac{m^2}{\square}\Bigr)^{-1}\theta(t - t_0) &=& \frac{(\omega_+ - 3iH)e^{i\omega_+(t - t_0)} - (\omega_- - 3iH)e^{i\omega_-(t - t_0)}}{\omega_+ - \omega_-}~~,~~t>t_0~~~\\
\label{coshdeg}
&=& e^{-\frac{3H}{2}(t-t_0)}\Bigl[cosh\Bigl(\frac{3H}{2}(t-t_0)\sqrt{1 - \frac{4m^2}{9H^2}}\Bigr)\\
\nonumber
 &+& \frac{1}{\sqrt{1 - \frac{4m^2}{9H^2}}}sinh\Bigl(\frac{3H}{2}(t-t_0)\sqrt{1 - \frac{4m^2}{9H^2}}\Bigr) \Bigr],
\end{eqnarray}

\noindent whose leading order behavior in the inverse filter scale $m$ is given by:

\eq{gm1}{\Bigl(1 + \frac{m^2}{\square}\Bigr)^{-1}\theta(t - t_0) =  \Bigl(1 + \frac{m^2}{9H^2}\Bigr)e^{-\frac{m^2}{3H}(t-t_0)} - \frac{m^2}{9H^2}e^{-3H(t-t_0)} \sim 1 - \frac{m^2}{2}(t-t_0)^2,}

\noindent which around $t_0$, mimics a Gaussian:

\eq{gaus}{\Bigl(1 + \frac{m^2}{\square}\Bigr)^{-1}\theta(t - t_0) = e^{-\frac{m^2}{2}(t-t_0)^2}.}

\noindent That is, the degravitation filter appears to transform the step function into a function that is zero before $t= t_0$, and decays exponentially afterwards with a characteristic time set by the inverse filter scale $t = m^{-1}$, which is typically larger than the present age of the universe.

In this way, we see that although degravitation kills any 'bare' cosmological constant term, energy densities which condense at some finite time into the history of the universe will require a timescale parametrically dependent on the inverse filter scale to degravitate. We apply these considerations to hybrid inflation \cite{linde}, which evidently relies on the spacetime zero mode of the inflaton potential to drive inflation. The potential for a generic hybrid inflation model is given by

\eq{hip}{V(\phi,\sigma) = \frac{1}{4\lambda}(M^2 - \lambda\sigma^2)^2 + \frac{m_\phi^2}{2}\phi^2  + \frac{g^2}{2}\phi^2\sigma^2,}

\noindent where the mass of $\phi$ is tuned such that it rolls very slowly. Given that the effective mass for the field $\sigma$ is equal to $-M^2 + g^2\phi^2$, for $\phi > \phi_c = M/g$, there is only one minima for the field $\sigma$ at $\sigma = 0$. In this regime, the zero mode piece of the potential dominates ($V = M^4/\lambda$) and drives inflation, which ends when $\phi$ rolls below $\phi_c$. Were we to apply the degravitation filter to (\ref{hip}), the zero mode piece of the potential would seem to be annihilated as evidenced from the expansion of the filter in (\ref{se}). However we note that one should really think of (\ref{hip}) as being multiplied by a step function in time which models the fact that the energy condenses at some finite time, or alternatively, models the beginning of the universe at some finite time in the past\footnote{To not do so would imply that this energy density has been around since the infinite past, by which time we expect degravitation to have naturally killed the constant part of the energy density. Hence in a non-eternal universe, we have to multiply all energy densities by a step function to model a universe with a beginning.}. Hence we conclude that degravitation leaves hybrid inflation unaffected provided it lasts for a duration less than the timescale associated with the filter scale, which it certainly does if $m^{-1} > H^{-1}_0$, where $H_0$ is the present day Hubble parameter.

We next consider the effect of degravitation on the following potential:

\eq{step}{V(t) = V_0 + \sum_{i=1}^{N}(V_i - V_{i-1})\theta(t - t_i),~~~~~t_{i-1} < t_i}

\noindent which models a series of potential drops from $V_{i-1}$ to $V_i$ at times $t_i$, i.e. $V(t) = V_k$ if $t_k < t < t_{k+1} $. Using the results from previous calculation, we find that the effect of the degravitation filter on the potential at a time $t_k < t < t_{k+1}$ is\footnote{Where we set $V_0 = 0$ to model a universe where a potential energy of $V_1 \neq 0$ appears at the beginning of the universe at time $t_1 = 0$, and successively cascades down a series of steps.}:

\begin{eqnarray}
\nonumber
\Bigl(1 + \frac{m^2}{\square}\Bigr) V(t) &=& \sum_{i=1}^k \frac{(V_i - V_{i-1})}{2}\Bigl[\Bigl(1 - \frac{1}{\sqrt{1 - \frac{4m^2}{9H^2}}}\Bigr)e^{-\frac{3H}{2}(t - t_i)(1 + \sqrt{1 - \frac{4m^2}{9H^2}})} \Bigr]~~~;~~~t_k < t < t_{k+1}\\
\label{vdeg}
&+& \sum_{i=1}^k \frac{(V_i - V_{i-1})}{2}\Bigl[\Bigl(1 + \frac{1}{\sqrt{1 - \frac{4m^2}{9H^2}}}\Bigr)e^{-\frac{3H}{2}(t - t_i)(1 - \sqrt{1 - \frac{4m^2}{9H^2}})} \Bigr],
\end{eqnarray}

\noindent which to second order in the inverse filter scale $m$, is given by

\eq{vdegs}{\Bigl(1 + \frac{m^2}{\square}\Bigr) V(t) = V_k -\frac{m^2}{2}\sum_{i=1}^k(V_i - V_{i-1})(t_i - t_{i-1})^2  ~~~;~~~t_k < t < t_{k+1}.}

\noindent Here we again see how degravitation leaves the potential unaffected for all time scales much smaller than the degravitation scale. Hence we see that old inflation, which can heuristically be modeled as such a single step function potential drop is unaffected provided its duration is microscopic compared to the degravitation scale. We note here that the degravitated potential at any given time is always going to be slightly larger than the undegravitated potential for any series of cascades, as degravitation carries a memory of the previous vacua (suppressed by powers of $m$) in the second term in the above. It is this observation that suggests that todays cosmological constant might be due to a similar such afterglow by considering the above in the context of some sort of potential (e.g. stringy) landscape. Supposing that our universe is presently in its true vacuum, but was once trapped in a series of false vacua, modeled by the transitions $V_1 \to V_2 \to ... \to V_N = 0$. We find that although the matter sector potential energy in the present epoch vanishes, degravitation implies an inherited memory of previous false vacua such that immediately after the final tunneling event, we still feel an remnant energy density:

\eq{ccd}{V_{rem} \leq \frac{m^2}{2}|\Delta V| \Delta T^2.}

\noindent In the above, $\Delta V$ is the total potential drop since the beginning of the universe and $\Delta T = t_0 - t_N$, where we note that the longer the duration of the period we are stuck in the false vacua, the stronger the afterglow we feel, as the non-locality of the degravitation filter has a longer duration source to act on. In the context of the Einstein equations $3H^2 = 8\pi G_N V_{rem}$, this implies an apparent cosmological constant

\eq{00ee}{ \Lambda = \frac{3H^2}{M^2_{pl}} = \frac{m^2}{M^2_{pl}}\frac{|\Delta V|\Delta T^2}{2M^2_{pl}},}

\noindent whose scale is an energy density in dimensionless units parametrically suppressed by the square of the filter scale in Planck units. Recalling that in requiring $m < H_0$, we see that

\eq{ccb}{\frac{m^2}{M^2_{pl}} < l^2_{pl}H^2_0 \sim \frac{l^2_{pl}}{R^2},}

\noindent where $R$ is the size of the universe today. Hence we see that although degravitation kills any spacetime zero mode (i.e. the bare cosmological constant), this calculation implies a remnant cosmological constant might still be observed today with a magnitude that is suppressed by the degravitation scale:

\eq{lobs}{\Lambda \sim \frac{l^2_{pl}}{L^2} \sim 10^{-120},}

\noindent if we take $L \sim H_0^{-1}$. In the following sections, we will rederive this result through reformulating the problem in a manner which admits generalization to spatially inhomogeneous sources. We will also uncover the result that in the scenario where we remain stuck in some false vacuum state, although degravitation causes the vacuum energy density to decay over time, it does so over a timescale that is typically too slow compared to the age of the universe. 

Before we continue however, we parenthetically note that all of the results and conclusions we have just discussed generalize rather straightforwardly to other values of $\alpha$ in (\ref{rg}). For example, having accounted for subtleties involving branch cuts in the integral (\ref{degs}), we find in the context of (\ref{dga}) that in place of $\omega_\pm$ as being given by (\ref{roots}), we have instead (for $0\leq \alpha < 1$)

\eq{rootsnew}{\omega_\pm = \frac{3iH}{2}\Bigl(1 \pm \sqrt{1 + \frac{4m^2}{9H^2}e^{i\pi(\frac{1}{1 - \alpha})}}\Bigr),}

\noindent with similarly straightforward generalizations for other results in the above. We thus continue to work with $\alpha = 0$ in the next section for quantitative convenience, with the results concerning values of $\alpha$ being similarly generalizable.

\section{An Afterglow Cosmological Constant?}

We can formally recast (\ref{mg}) as

\eq{mg2}{\frac{8\pi G_N}{1 + \frac{m^2}{\square}} = \frac{8\pi G_N\square}{\square + m^2} = 8\pi G_N\square (\square + m^2)^{-1},}

\noindent so that its action on any source takes the form

\eq{dgs2}{8\pi G_N\Bigl(\frac{\square}{m^2}\Bigr)\rho (x) = 8\pi G_N \square \int d^4x'\sqrt{-g(x')} G(x,x')\rho(x'),}

\noindent where $G(x,x')$ satisfies

\eq{gfe}{(\square_x + m^2)G(x,x') = \frac{\delta(x,x')}{\sqrt{-g(x')}}.}

\noindent Acting through with $\square$ in (\ref{dgs2}) and using (\ref{gfe}), we obtain the result

\eq{rdg}{\rho_{degrav}(x) = \rho(x) - m^2 \int d^4x'\sqrt{-g(x')} G(x,x')\rho(x'),}

\noindent which makes manifest the causality (and non-locality) of degravitation for the appropriate choice for the Greens function in the above. Furthermore, upon integrating (\ref{gfe}) we find
\begin{equation}
\label{gfeint}
1 = m^2\int d^4x \sqrt{-g(x)}G(x,y) + \int_{\partial \mathcal M}d^3\xi~ n^a\partial_aG(\xi,y),\\
\end{equation}

\noindent where we note that the surface integral vanishes at coordinate infinity by virtue of the finite mass scale $m$ (which results in exponentially damped asymptotic behavior of the Green's function). Hence we find that the Green's function is normalized as $1 = m^2\int d^4x' \sqrt{-g(x')}G(x',x)$\footnote{The integral is independent of the argument $x$, which we pick to be the origin in the following}, which allows us to recast (\ref{rdg}) as

\eq{m2}{\rho_{degrav}(x) = \rho(x) - \frac{\int d^4x'\sqrt{-g(x')} G(x,x')\rho(x')}{\int d^4x' \sqrt{-g(x')}G(0,x')}.}

\noindent If we assume that we are in a de Sitter background, then we know that the retarded Greens function is\cite{bd}

\eq{gf}{G(x,x') = \theta(x_0 - x_0') \mathcal Re ~i \frac{2H^2}{16\pi^2}\Gamma\Bigl(\frac{3}{2} + \nu\Bigr)\Gamma\Bigl(\frac{3}{2} - \nu\Bigr) F\Bigl(\frac{3}{2} + \nu, \frac{3}{2} - \nu,2, \frac{1+z + i\epsilon}{2}\Bigr),}

\noindent with $z$ being the de Sitter invariant geodesic distance \footnote{See \cite{sv} for this and other aspects of de Sitter space physics used here.} between $x$ and $x'$ and $\nu^2 = 9/4\sqrt{1- 4m^2/9H^2}$. This suggests an interpretation of (\ref{m2}) as the weighted average:

\eq{avg}{\rho_{degrav}(x) = \rho(x) - \langle \rho \rangle_x,}

\noindent with the average defined by:

\eq{avdef}{\langle f \rangle_x := \frac{\int d^4x'\sqrt{-g(x')} G(x,x')f(x')}{\int d^4x' \sqrt{-g(x')}G(0,x')}.}

\noindent We immediately see how a spatial zero mode (bare cosmological constant) is degravitated, as

\eq{ldhg}{\Lambda_{degrav} = \Lambda (1 - \langle 1\rangle) = 0.}

\noindent Consider now a source corresponding to a bubble of one vacuum nucleating in another starting at time $t = t_0$ at the origin:

\eq{nuc}{V(x) = V_0 + (V_f - V_0)\theta[t-t_0]\theta[z(x,0) - 1],}

\noindent where we see that the product of step functions effects a potential of $V_0$ everywhere except inside the future lightcone of the nucleation event, where it is $V_f$. Consider the effect of the degravitation filter on the step function through (\ref{avg}) and (\ref{avdef}):

\eq{dsf}{\theta[t-t_0]\theta[z - 1]_{degrav}  = \theta[t-t_0]\theta[z - 1] - \frac{\int_\blacklozenge \sqrt{-g} G(x)}{\int_\blacktriangle \sqrt{-g}G},}

\noindent where $z$ is the de Sitter invariant geodesic distance to the nucleation event ($z = 1$ implies zero or null separation), $\int_\blacklozenge$ indicates an integral over the causal diamond bounded by $x$ in the present, and the nucleation event in the past and $\int_\blacktriangle$ indicates an integral over the entire past light cone of any observer at the origin (see footnote associated with (\ref{m2})). For $t < 0$ there is no causal diamond between $x$ and the nucleation event as it is yet to happen (the theta function vanishes in the integrand), so we get:

\eq{tdgq}{\theta[t-t_0]\theta[z - 1]_{degrav}  = \theta[t-t_0]\theta[z - 1]  = 0 ~~~~~ t < t_0,}

\noindent which is exactly as we had obtained in our previous example (\ref{dgb}). For $t > t_0$, the integral over the causal diamond commences at zero and asymptotes towards the integral over the entire past light cone of the origin event ($\int_\blacklozenge G \to \int_\blacktriangle G$). Hence

\eq{dec}{\theta[t-t_0]\theta[z - 1]_{degrav} = 1 - f(t)~~;~~ f(0) = 0~~\to~~ f(\infty) = 1.}

\noindent Since the function $G(x,x')$ describes the propagation of a massive particle, we expect it to be an exponentially decaying function of the geodesic distance between $x$ and $x'$, with a characteristic scale set by $m^2$. To see this (and to reproduce the results of the last section), we work in co-ordinates where

\eq{crgc}{ds^2 = \frac{1}{H^2\eta^2}(d\eta^2 -dx_idx^i),}

\noindent and consider for simplicity, a homogeneous source as in the previous section. In this case, the relevant source is $\theta(x) = \theta(\eta - \eta_0)$. In this case, (\ref{rdg}) becomes

\eq{thetadeg}{\theta_{degrav}(\eta - \eta_i) - m^2\int d\eta'd^3x' \sqrt{-g(x')}G(x,x')\theta(\eta' - \eta_i),}

\noindent with $G(x,x')$ given as in (\ref{gf}). Since our source is now spatially homogeneous, we can integrate the Green's function over the spatial coordinates to give us \cite{spec}

\begin{eqnarray}
\label{spint}\int d^3x'\sqrt{-g(\eta')}G(x,x') &=& \frac{\theta(\eta - \eta')(-\eta)^{3/2}}{H^2\nu (-\eta')^{5/2}}sinh[\nu ln (\eta/\eta')]\\
&=& \nonumber \frac{\theta(\eta - \eta')}{2H^2\nu}\Bigl( \frac{\eta^{3/2 - \nu}}{\eta'^{5/2 - \nu}} - \frac{\eta^{3/2 + \nu}}{\eta'^{5/2 + \nu}} \Bigr),
\end{eqnarray}

Hence the integral we have to perform is:

\begin{eqnarray}
\label{intperf} \int_\blacktriangle d^4x' \sqrt{-g(x')} G(m^2;x,x') &=& \frac{1}{2H^2\nu}\int^\eta_{\eta_i} d\eta' \Bigl( \frac{\eta^{3/2 - \nu}}{\eta'^{5/2 - \nu}} - \frac{\eta^{3/2 + \nu}}{\eta'^{5/2 + \nu}} \Bigr)\\
\nonumber
&=& \frac{1}{m^2} - \frac{1}{2\nu m^2}\Bigl[ (\nu + 3/2)(\eta/\eta_i)^{3/2 - \nu} + (\nu - 3/2)(\eta/\eta_i)^{3/2 + \nu}  \Bigr],
\end{eqnarray}

\noindent from which we see that if we were to push the initial time $\eta_i \to -\infty$, the integral asymptotes to $1/m^2$, as implied by (\ref{gfeint}). When we rewrite conformal time in terms of cosmological time:

\eq{ctct}{\eta = -\frac{e^{-Ht}}{H},}

\noindent we see that (\ref{intperf}) becomes:

\eq{intperf2}{\frac{1}{m^2}\Bigl[ 1 - e^{-3H(t-t_i)/2}\Bigl( cosh[\nu H(t-t_i)] + \frac{3}{2\nu } sinh[\nu H(t-t_i)]\Bigr) \Bigr],}

\noindent so that the degravitated step function becomes:

\begin{eqnarray}
\label{degpot}
\theta(t - t_i)_{degrav} &=& 0~~; t < t_i\\
\nonumber &=&  e^{-3H(t-t_i)/2}\Bigl( cosh[\nu H(t-t_i)] + \frac{3}{2\nu } sinh[\nu H(t-t_i)]\Bigr)\\
\nonumber &=& e^{-3H(t-t_i)/2}\Bigl( cosh[\sqrt{1 - \frac{4m^2}{9H^2}}3H(t-t_i)/2]\\
\nonumber
&+& \frac{1}{\sqrt{1 - \frac{4m^2}{9H^2}}} sinh[\sqrt{1 - \frac{4m^2}{9H^2}}3H(t-t_i)/2]\Bigr) ;~~ t > t_i,
\end{eqnarray}

\noindent which is exactly as in (\ref{dga}). From this, we can again follow the logic which led up to (\ref{ccd}) and (\ref{lobs}), allowing us the possibility that although the degravitation mechanism has rendered the spacetime zero mode of the energy momentum tensor gravitationally null, it does so in a way that preserves inflation, and in such a way that it can account for the presently observed value of the cosmological constant as a memory of previous energy densities. The scales of our problem allow us the possibility that the value for the cosmological constant

\eq{lobs2}{\Lambda \sim 10^{-120}}

\noindent is set through the hierarchy of scales that is the filter scale in Planck units (as in (\ref{00ee})). It might appear to us at this point that we have replaced one tuning for another, in that now the smallness of the cosmological constant depends on the extreme smallness of the filter scale (albeit a smallness that is forced on us by the consistency of GR at most scales accessible to our observations). However in the context of the filter arising from a small graviton mass, this tuning is technically natural, in that the graviton mass is radiatively stable. The reason for this can be understood from the fact that the graviton mass is protected quantum mechanically (e.g. against gauge field loops) by the gravitational Ward identities, hence any small mass causes the quantum mechanical corrections to the graviton mass to be proportional to the mass itself, and hence can be neglected.

We now turn our attention to the scenario that our universe is still stuck in some meta stable vacuum state (as implied by the notion that we live in potential landscape). We wish to explore whether or not degravitation can degravitate this false vacuum energy to a reasonably small value over the age of the universe, and whether or not it does so in a way that addresses the coincidence problem. As we shall see, the class of degravitation models \cite{dhk} that we study here act over time scales that are far too slow to degravitate false vacuum energy densities in a way that addresses the coincidence problem \footnote{However it might be that there are models that act over much faster time scales \cite{addg} which might work in this regard, we postpone this to a future study.}.

\section{The Coincidence Problem}

To offer us further perspective on the degravitation mechanism, and to study degravitation in other contexts, we reformulate the problem in yet another manner. We begin by acting on both sides of (\ref{mg}) with appropriate powers of $\square$ and $\square + m^2$, to rewrite the modified Einstein's equations as

\eq{rmee}{8\pi G_N T^\mu_\nu = G^\mu_\nu + m^2 \square^{-1} G^\mu_\nu,}

\noindent whence the $00$ equation becomes

\eq{inb}{8\pi G_N \rho(x) = 3H^2(x) + m^2\int \sqrt{-g(x)} G^0(x,x')3H^2(x'),}

\noindent and where $G^0(x,x')$ satisfies

\eq{mgf}{\square_x G^0(x,x') = \frac{\delta(x,x')}{\sqrt{-g(x)}}.}

\noindent Rewriting the above as

\eq{rew}{3H^2(x') = 8\pi G_N\rho(x') - m^2 \int \sqrt{-g(x'')}dx''G^0(x',x'')3H^2(x''),}

\noindent and substituting in the integrand in (\ref{inb}) and iterating, for a step function potential drop $\rho =\Delta V\theta[t-t_0]\theta[z - 1] := \theta $, we arrive at the expression

\eq{degh}{3H^2 = \frac{\Delta V}{M^2_{pl}}\Bigl[\theta - m^2(G^0,\theta) + m^4( G^0,( G^0,\theta)) -  m^6( G^0,( G^0,( G^0,\theta))) + ...  \Bigr],}

\noindent where $G^0$ is the retarded Greens function for a minimally coupled massless scalar field in the fully degravitated spacetime (for its expression in pure de Sitter space, see for example \cite{kg}), and $(G^0,\theta) := \int_{\blacklozenge}G^0$ is the integral over the domain $\blacklozenge$, again defined as the causal diamond bounded by the nucleation event and the observation event. Clearly the right side of the above vanishes for $t < t_0$. For $t >  t_0$, realizing the ordered nature of the multiple integrals in the above, {\it for strictly homogeneous sources} (i.e. ones which are only functions of time) we can re-express (\ref{degh}) as

\eq{ressum}{3H^2 = \frac{\Delta V}{M^2_{pl}}\Bigl[1 - m^2\Delta + m^4\frac{\Delta^2}{2!} -  m^6\frac{\Delta^3}{3!} + ...  \Bigr],}

\noindent where $\Delta$ is given by

\eq{del}{\Delta = \int_\blacktriangle G^0,}

\noindent and $\blacktriangle$ denotes the backward lightcone of the observation event. This expression easily resums to the following expression for the degravitated step function\footnote{the corresponding expression for a nucleating bubble  (which is manifestly inhomogeneous) is not so easily resummable.}

\eq{degstep2}{\theta(t - t_i)_{degrav} = e^{-m^2\int_\blacktriangle G^0}~~;~~ t> t_i.}

\noindent In the context of (\ref{ressum}), this implies the rather complicated integral equation

\eq{eqcomp}{3H^2 = \frac{\Delta V}{M^2_{pl}}e^{-m^2\int_\blacktriangle G^0[H]},}

\noindent where the implicit dependence on $H$ in the integrand is highlighted. This equation is exact, and in principle (though not in practice) solvable. We can make progress by invoking an adiabatic approximation, namely that we take the scale that sets spacetime curvature $H$ to be a very slowly varying function of co-ordinate time. This implies at the very least that $m^2 << H^2$. For convenience, we also consider the modified filter function\footnote{In general, one might even expect a filter function of the form (\ref{mff}) as powers of $\square$ and multiples of the identity operator mix readily under renormalization group transformations.}

\eq{mff}{\frac{8\pi G_N}{1 + \frac{m^2}{\square + \mu^2}},}

\noindent where $\mu$ is some other IR length scale that is taken to be much less than $m$, and take the appropriate limit at the end of our calculations. In this case we find that (\ref{mff}) results in the expression

\eq{eqcomp2}{\theta(t - t_i)_{degrav} = e^{-m^2\int^t_{t_i}\int d^3x G(x,x')},}

\noindent with $G(x,x')$ given by (\ref{gf}), but with $\kappa$ in place of $\nu$, with $\kappa$ given by

\eq{newn}{\kappa^2 = 9/4 - \mu^2/H^2.}

\noindent As in the previous section, we can evaluate the integrand in the above to yield

\eq{eqdf}{\theta(t - t_i)_{degrav} = e^{-\frac{m^2}{\mu^2}\Bigl(1 - \frac{1}{2\kappa}\{(\kappa + 3/2)exp[-H(3/2 - \kappa)(t-t_i)] + (\kappa - 3/2)exp[-H(\kappa + 3/2)(t-t_i)]\}\Bigr)},}

\noindent which in the limit $\mu \to 0$ evaluates to:

\eq{gausss}{\theta(t - t_i)_{degrav} = e^{-\frac{m^2}{2}(t-t_i)^2},}

\noindent exactly as in (\ref{gaus}). Hence we can infer in a slightly different context, that the relevant timescale for degravitation to effect itself is set by the inverse filter scale $\tau = m^{-1}$. We thus reason that we were to require degravitation to degravitate Planck or GUT scale energy densities (as one might expect if we were stuck in some metastable vacuum state in the landscape), down to the presently observed value, we then require that a time of at least an order of magnitude larger than $m^{-1}$ to have elapsed. That is, from (\ref{gausss}) we can estimate that we require a time interval of $\Delta t \sim m^{-1}\sqrt{276}$ to elapse\footnote{The factor 276 comes from the fact that $e^{-276} = 10^{-120}$.}. Typically, we require $m^{-1} > H_0^{-1}$, where $H_0$ is the presently observed Hubble scale. Hence we find that unless the universe began much earlier than we infer from current measurements of $H_0$ (as is entirely likely if we live in an eternally inflating landscape), that degravitation is unlikely to have degravitated primordial string or Planck scale energy densities sufficiently.

However, the main observation of this section, is that regardless of whether or not degravitation has had enough time to act in order to degravitate potentially high scale primordial energy densities, if it is to satisfactorily address the coincidence problem, it has to act on a timescale comparable to the {\it present} Hubble scale:

\eq{phs}{\Delta t\sim H^{-1}_0,}

\noindent which we have already demanded not be the case so as not to conflict with other cosmological observations. Hence it appears as if the degravitation filter (\ref{rg}) acts too slowly to be of any help in addressing the coincidence problem without additional tuning. To conclude this report, we rework the calculation for how the degravitation filter degravitates a step function source for a different functional form for the filter function. Consider the sample filter function:

\begin{eqnarray}
8\pi G_N &\to& 8\pi G_N \bigl[ln\Bigl(\frac{e\square + \mu^2}{\mu^2}\Bigr) - ln\Bigl( \frac{\square + \mu^2}{\mu^2}\Bigr) \Bigr]\\
&=& 8\pi G_N \bigl[1 + ln\Bigl(\frac{\square + \mu^2/e}{\square + \mu^2}\Bigr) \Bigr]
\end{eqnarray}
which is easily checked to yield a degravitating filter function. We chose this functional form only for illustrative purposes. Using the representation\cite{barv1}\cite{spectral}
\eq{repsln}{ln[\square/\mu^2] = \int^\infty_0 d\kappa^2 \Bigl( \frac{1}{\mu^2 + \kappa^2} - \frac{1}{\square + \kappa^2}\Bigr),}    
we find that the action of such a filter function on a step function potential would result in
\begin{eqnarray}
\nonumber
\theta_{degrav}(t-t_0) &=& Chi\Bigl(\frac{3H\Delta t}{2}[1 - \sqrt{1 - 4\mu^2/9H^2}]\Bigr) - Shi\Bigl(\frac{3H\Delta t}{2}[1 - \sqrt{1 - 4\mu^2/9H^2}]\Bigr)\\ \nonumber
&+& Chi\Bigl(\frac{3H\Delta t}{2}[1 + \sqrt{1 - 4\mu^2/9H^2}]\Bigr) - Shi\Bigl(\frac{3H\Delta t}{2}[1 - \sqrt{1 + 4\mu^2/9H^2}]\Bigr)\\
&-& [...]_{\mu^2 \to \mu^2/e}
\end{eqnarray}
where $Chi(x)$ and $Shi(x)$ are the hyperbolic cosine integral and hyperbolic sine integral functions respectively. We compare this expression to (\ref{coshdeg}) to infer the same characteristic dependence on the inverse filter scale, such that the timescale associated with degravitation is given by $\tau = \mu^{-1}$. In this way, we infer that the conclusions we drew in the previous sections would be true of this filter function as well, implying that degravitation appears to be somewhat insensitive to the precise functional form of the filter function. We now offer our concluding thoughts.

\section{Conclusions}

In this report, we have seen in detail how degravitation works in annihilating the bare cosmological constant whilst preserving inflation in all of its forms. We have also demonstrated how degravitation inherits a memory of previous energy densities in such a way that even if our universe were to exist in its true vacuum state today, degravitation would imply an afterglow cosmological constant which can naturally be arranged to mimic the dark energy that we infer today. The key physics of this observation is that such an afterglow is suppressed by the square of the inverse filter scale in Planck units\footnote{In the context of massive gravity, it is interesting to note how our results appear as the reverse interpretation of the bound $m^2 \geq \Lambda/3$ derived by Deser and Waldron concerning consistent values for the mass of the graviton in de Sitter backgrounds.} $m^2/M^2_{pl} = l^2_{pl}/L^2 < 10^{-120}$. We then showed that if we exist in a universe which is still trapped in some false vacuum state, degravitation can degravitate the energy of such a false vacuum into a remnant energy density of the order $\Lambda \sim 10^{-120}$, however this typically occurs over timescales at least an order of magnitude larger than the age of the universe. In this way, although degravitation answers why the cosmological constant is not large (the bare cosmological constant problem), as well as why it is not zero (that the cosmological constant might be the dark energy that we observe today)\cite{pol}, it does not satisfactorily explain the coincidence problem, or why it only begins to dominate now (although this is not to say that other models of degravitation may not have something to say about this). However, any mechanism that contains hints of being able to address all three aspects of the cosmological constant problem if further developed (while preserving the successes of inflationary cosmology) certainly warrants closer attention, and we hope that the findings of this report will motivate further investigation into this promising paradigm.

\section{Acknowledgements}
I am grateful to Gia Dvali and the CCPP at NYU for hospitality during the time in which this work was initiated, and for illuminating discussions on the sidelines of various conferences throughout the year. Thanks to Justin Khoury and the Perimeter Institute for the invitation to attend a stimulating conference and many useful exchanges. Thanks to Gregory Gabadadze for many useful discussions, perspectives and a copy of the preprint \cite{cern}. I remain indebted to Robert Brandenberger for his continued support and frequent dialogue. Special thanks to Zoe Greenberg, whose presence in my life has left a wake of inspiration that persists. The portion of this work undertaken whilst still at McGill University was supported in part by an NSERC discovery grant. This work is supported at the Humboldt University by funds from project B5 of the Sonderforschungsbereich 647 (Raum Zeit Materie) grant, for which I am grateful to Alan Rendall at Albert Einstein Institute and Jan Plefka at the Humboldt University, whom I also thank for his continued support and encouragement.

\end{document}